# Unique Phase Transition on Spin-2 Triangular Lattice of $Ag_2MnO_2$

Hiroyuki Yoshida, Sascha Ahlert,[1] Martin Jansen,[1] Yoshihiko Okamoto, Jun-Ichi Yamaura, and

Zenji Hiroi[*]

*Institute for Solid State Physics, University of Tokyo, Kashiwa, Chiba 277-8581, Japan*

[1]*Max-Planck-Institut für Festköperforschung, Heisenbergstrasse 1, 70569 Stuttgart, Germany*

$Ag_2MnO_2$ is studied as a possible candidate compound for an antiferromagnetic *XY* spin model on a triangular lattice. In spite of the large Curie-Weiss temperature of -430 K found in magnetic susceptibility, $Mn^{3+}$ spins with $S = 2$ do not undergo a conventional long-range order down to 2 K probably owing to the geometrical frustration and two dimensionality in the system. Instead, a unique phase transition is found at 80 K, where specific heat exhibits a clear sign of a second-order phase transition, while magnetic susceptibility changes smoothly without a distinct anomaly. We think that this transition is related to the chirality degree of freedom associated with a short-range order, which has been expected for the classical *XY* spin model on a triangular lattice. On further cooling, spin-glass-like behavior is observed below 22 K, possibly corresponding to a quasi-long-range order.

*E-mail address: hiroi@issp.u-tokyo.ac.jp

## 1. Introduction

Frustration is a key ingredient in the recipe for unconventional phenomena or novel phases in spin systems on various triangular lattices.[1, 2] When spin-carrying atoms set in triangular arrays with spins interacting antiferromagnetically with their neighbors, attempts of the spins to order themselves are frustrated, and an exotic state appears instead of a conventional long-range order (LRO). Another important factor in line with recent trends is quantum fluctuation, which is expected for spins of small quantum number such as $S = 1/2$; this also suppresses LRO. It is predicted theoretically that quantum spins on a triangular lattice will not freeze into a fixed configuration, even at the lowest temperature, resulting in a unique state called the spin liquid.[3, 4]

Lured by the prospect of finding interesting physics, experimentalists have been searching for real materials that embody such behavior.[4] Several compounds have been studied thus far; for example, herbertsmithite characterized by the ideal formula $ZnCu_3(OH)_6Cl_2$ and volborthite by $Cu_3V_2O_7(OH)_2·2H_2O$ have been examined as candidates for the $S = 1/2$ kagome lattice,[5-8] and the organic compound $\kappa$-$(BEDT-TTF)_2Cu_2(CN)_3$ for the $S = 1/2$ triangular lattice.[9] $NiGa_2S_4$ is another compound with a triangular lattice but with $S = 1$.[10] All of them exhibit neither LRO at low temperatures, e.g., ~50 mK, nor the spin gap expected for a short-range resonating-valence-bond state. The ground state of these compounds is likely a quantum-disordered state or a nonmagnetic Mott insulator.[9, 11]

On the other hand, it is well known that frustration gives rise to more pronounced effects on the properties of classical spin systems, in which the anistropy of spins or interactions predominates the ground states and phase transitions. There is no order for the antiferromagnetic (AF) Ising model on the triangular lattice, while the so-called 120˚ structure is stabilized marginally for the *XY* and Heisenberg model at $T$ = 0 K.[12] The presence of order at $T$ = 0 K means that a certain phase transition to a disordered state will occur at an elevated temperature. Because of two dimensionality, however, conventional LRO may not be allowed at a finite temperature, and a sort of short-range order (SRO) involving pairs of defects in the 120˚ structure called vortices is realized.[12] Thus, the transition is classified as the Kosterlitz-Thouless (KT) transition, where pairs of vortices are dissociated.[13]

The AF *XY* spin model on a triangular lattice has been studied theoretically in the past two decades.[14-16] In 1984, Miyashita and Shiba (MS) studied the AF *XY* model on a triangular lattice by Monte Carlo simulations and found two phase transitions:[14] one is the above-mentioned KT transition at $k_BT_{KT} = 0.502J$ and the other is a novel order-disorder transition associated with the chirality degree of freedom at $k_BT_{ch} = 0.513J$, where $k_B$ is the Boltzmann constant and $J$ is the AF interaction between nearest-neighbor spins. Chirality is defined as the sign of rotation of three spins forming an angle of 120˚ to each other at vertices of each elementary triangle: the 120˚ structure is described as a staggered order of chirality. Interestingly, MS found that specific heat shows a logarithmic divergence at $T_{ch}$, while magnetic susceptibility is almost temperature independent at around $T_{ch}$, even though the transition is associated with the ordering of spins. Moreover, they suggested the possible existence of an anomalous intermediate state between $T_{KT}$ and $T_{ch}$, where translational spin-order vanishes, while chirality order remains. However, because above two transitions are expected to occur at very close temperatures, it is not clear whether they are independent of each other or inseparable.[17]

A frequent obstacle for experimentalists is the difficulty in finding an appropriate material in which such an interesting theoretical prediction can be tested: a real material is not purely two-dimensional (2D), or the anisotropy of spins or interactions is always in between two extremes. The most well studied are the so-called stacked-triangular antiferromagnets like $CsCoCl_3$ for the Ising model, $CsVCl_3$ for the Heisenberg model and $RbFeCl_3$ for the *XY* model.[12] All of them exhibit LRO in the 120˚ structure below high critical temperatures owing to their three dimensionality. Rare-earth manganites $RMnO_3$ ($R$ = Y, Lu and Sc) have been investigated recently as a candidate AF *XY* spin model on a triangular lattice.[18-20] They crystallize in a hexagonal structure with $Mn^{3+}$ ions with $S = 2$ arranged in a distorted triangular lattice or a coupled-trimer lattice. Mn spins undergo LRO in the 120˚ structure: for example, in the case of $YMnO_3$, the ordering temperature $T_N$ is 70 K, which is much lower than the Curie-Weiss temperature $\Theta_{CW}$ = -705 K. Curiously, inelastic neutron scattering experiments on $YMnO_3$ revealed that unusual 2D AF spin fluctuations survive even below $T_N$, which bears characteristics of the KT phase intrinsic to the 2D *XY* spin systems.[20]

In this paper, we present the novel compound $Ag_2MnO_2$ as a possible candidate AF *XY* spin model on a triangular lattice. We found a unique phase transition at $T_c$ = 80 K, where specific heat exhibits a clear sign of second-order transition, while magnetic susceptibility changes smoothly, similar to the prediction by MS for chirality transition.[14]

$Ag_2MnO_2$, discovered in 1963, was first regarded to contain divalent manganese.[21] In the course of recent reinvestigations, its similarity to $Ag_2NiO_2$ has been noted.[22] $Ag_2NiO_2$ crystallizes in a trigonal structure of the space group $R$-$3m$, in which $[Ag_2]^+$ and $[NiO_2]^-$ layers stack alternatingly,[23, 24] as depicted in the inset of Fig. 1. The $[Ag_2]^+$ layer gives rise to a metallic conductivity ascribed to a quarter-filled Ag 5$s$ band, and the $[NiO_2]^-$ layer presents an $S$ = 1/2 triangular lattice made of $Ni^{3+}$ ions with a 3$d^7$ configuration in the low-spin state. $Ag_2NiO_2$ exhibits a structural transition at $T_s$ = 260 K probably ascribed to orbital ordering and AF order at $T_N$ = 56 K.[25, 26] In contrast, the crystal structure of $Ag_2MnO_2$ has not yet been determined thoroughly, although it is supposed to be similar to $Ag_2NiO_2$. There must be a structural phase shift between the layers, resulting in a small deviation from the trigonal symmetry. Thus, we assume for $Ag_2MnO_2$ a slightly distorted triangular lattice made of $Mn^{3+}$ ions that carry $S = 2$ in the high-spin configuration for the 3$d^4$ system.

## 2. Experimental

Polycrystalline samples of $Ag_2MnO_2$ were prepared by a solid state reaction between a fine powder of Ag metal and $MnO_2$. They were mixed in a stoichiometric ratio in an agate mortar, pressed into a pellet and calcined at 650˚C in an oxygen flow for 24 hs. The obtained pellet was crushed in an agate mortar and calcined again at 750˚C several times to achieve phase equilibrium. The product is brown in color with metallic luster and easy crush into thin fine particles, reflecting the inherent two dimensionality of the crystal structure.

Structural analyses were performed by means of electron and X-ray diffraction (XRD). An electron diffraction pattern taken with an incident beam perpendicular to a platelike crystal showed nearly hexagonal symmetry with in-plane lattice constants similar to that for $Ag_2NiO_2$. The powder XRD pattern of the product showed a strong tendency to a preferred orientation, which prevented us from performing a complete Rietveld analysis. However, we could determine the possible crystal system and lattice constants: monoclinic, $a$ = 5.178 Å, $b$ = 2.875 Å, $c$ = 8.815 Å, $\beta$ = 102.3˚. With these parameters, all the peaks observed in the XRD pattern could be indexed, indicating a phase-pure sample. A complete structural refinement is still in progress.

dc and ac magnetic susceptibilities were measured on polycrystalline samples in a Quantum Design magnetic property measurement system in the temperature range between 2 and 350 K. Resistivity measurement was performed by the four-probe method in a Quantum Design physical property measurement system (PPMS). Specific heat meaurement was carried out by the heat-relaxation method in the same PPMS.

## 3. Results

*3.1 Electrical properties*

$Ag_2MnO_2$ is a good metal the same as $Ag_2NiO_2$. Figure 1 shows the temperature dependence of resistivity measured on cooling and then on heating for $Ag_2MnO_2$. It shows metallic behavior with small $\rho$ values in spite of the fact that the measurement was carried out on a pressed pellet. No anomalies due to phase transition or superconductivity are observed down to 2 K. This is in contrast to the case of $Ag_2NiO_2$, where two anomalies at $T_s$ and $T_N$ exist.[25] The resistivity is proportional to $T^2$ at low temperatures below ~50 K, $\rho = \rho_0 + AT^2$, as indicated by the $T^2$ plot shown in the inset. The coefficient $A$ and the residual resistivity $\rho_0$ are determined to be $A$ = 2.86 × $10^{-4}$ $\mu\Omega$ cm $K^{-2}$ and $\rho_0$ = 1.42 $\mu\Omega$ cm. The $T^2$ dependence may indicate that elec-tron-electron scattering is predominant at low temperatures rather than electron-phonon scattering.

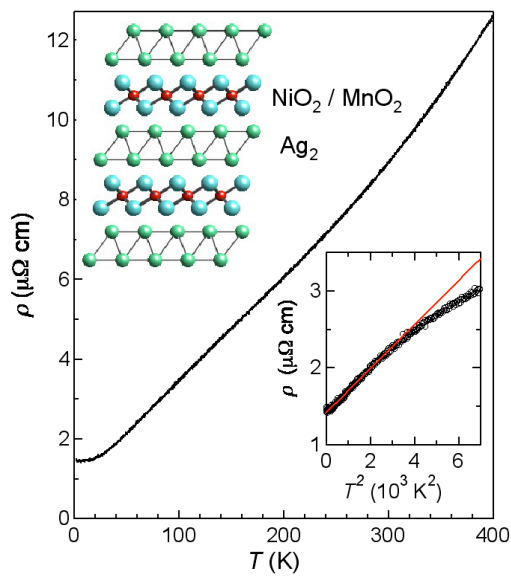

Fig. 1. Resistivity of $Ag_2MnO_2$ measured on cooling and heating. The lower inset shows a $T^2$ plot at low temperatures with a linear fit presented by a solid line. The upper inset shows the layered crystal structure of $Ag_2NiO_2$. That of $Ag_2MnO_2$ is supposed to be basically similar to this, consisting of conducting $Ag_2$ layers and magnetic $MnO_2$ layers containing a triangular lattice made of $Mn^{3+}$ ions with $S = 2$.

A $T$ linear term in specific heat, which is characteristic of a metal, is observed in the $C/T$ versus $T^2$ plot in Fig. 2. From the linear relation expressed as $C/T = \gamma + \beta T^2$, we obtain $\gamma = 20.6$ mJ K$^{-2}$ mol$^{-1}$ and $\beta = 0.744$ mJ K$^{-4}$ mol$^{-1}$. This $\gamma$ is close to that of Ag$_2$NiO$_2$ (18.8 mJ K$^{-2}$ mol$^{-1}$)[25] and is notably large compared with those of other subvalent silver compounds with a broad Ag 5$s$ band. For example, Ag$_2$F and Ag$_7$O$_8$NO$_3$, both of which are superconductors with $T_c$ = 66 mK and 1.0 K, respectively, have $\gamma = 0.62$ mJ K$^{-2}$ mol$^{-1}$ and 7.28 mJ K$^{-2}$ mol$^{-1}$ (about 1 mJ K$^{-2}$ for 1 mol of Ag), respectively.[27, 28] Therefore, a large increase in $\gamma$ occurs in Ag$_2$MnO$_2$, as in Ag$_2$NiO$_2$. The Debye temperature $\Theta_D$ is deduced from $\beta$ to be 236 K, which is nearly equal to that of Ag$_2$NiO$_2$, $\Theta_D = 239$ K.[25] This reflects the basic similarity between the crystal structures of the two compounds.

The coefficient $A$ of the $T^2$ term in resistivity and $\gamma$ from specific heat yield the Kadowaki-Woods ratio of $A/\gamma^2 = 7.2 \times 10^{-7}$, close to the universal value of $4 \times 10^{-7}$ for a metal with weak electron correlations. This may be consistent with the fact that metallic conduction occurs predominantly in a wide Ag 5$s$ band. The interplay between conduction electrons in the Ag$_2$ layers and magnetic moments in the MnO$_2$ layers can be expected and would be clarified in a future study using a single crystal.

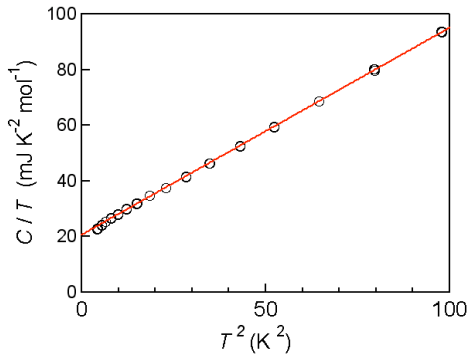

Fig. 2. Specific heat divided by temperature plotted against $T^2$. The solid line is a linear fit to the data, yielding $\gamma = 20.6$ mJ K$^{-2}$ mol$^{-1}$.

*3.2 Magnetic properties*

Figure 3 shows the magnetic susceptibility $\chi$ and its inverse in a wide temperature range. $\chi$ increases gradually on cooling from 350 to ~100 K, almost perfectly following the Curie-Weiss law, $\chi_{CW} = C_C/(T - \Theta_{CW})$, as evidenced by the linear $T$ dependence in the inverse $\chi$ plot. From the linear fit to the inverse $\chi$ data, one obtains a Curie constant $C_C$ of 2.70 cm$^3$ K$^{-1}$ mol$^{-1}$ and a Curie-Weiss temperature $\Theta_{CW}$ of -370 K. However, it is noticed in more careful examinations that the inverse $\chi$ shows a small concave upward curvature, which implies that there is a small temperature-independent contribution. Thus, we alternatively fitted the $\chi$ data assuming that $\chi = \chi_0 + \chi_{CW}$ for 100 K < $T$ < 350 K, and obtained that $\chi_0 = -1.2(5) \times 10^{-4}$ cm$^3$ mol$^{-1}$, $C_C = 3.02(7)$ cm$^3$ K$^{-1}$ mol$^{-1}$ and $\Theta_{CW} = -429(7)$ K. The $\chi_0$ term is negative but is very small compared with the total spin susceptibility, e.g., $\chi = 4 \times 10^{-3}$ cm$^3$ mol$^{-1}$ at 300 K.

In general, a temperature-independent term in $\chi$ is assumed to contain three contributions: diamagnetism from the core electrons of atoms ($\chi_c$), Pauli paramagnetism from conduction electrons ($\chi_P$), and orbital magnetism from 3$d$ electrons ($\chi_{orb}$). $\chi_c$ is evaluated to be $-0.8 \times 10^{-4}$ cm$^3$ mol$^{-1}$ from the literature, and the $\chi_P$ is estimated to be $1.8 \times 10^{-4}$ cm$^3$ mol$^{-1}$ from $\gamma$ assuming that the Wilson ratio is unity. Then, these values yield the remaining term $\chi_{orb} = -2.2 \times 10^{-4}$ cm$^3$ mol$^{-1}$. However, such a negative value for $\chi_{orb}$ is unrealistic. The reason for the negative $\chi_0$ term is not clear. Anyway, $\chi_0$ is so small compared with $\chi_{CW}$ that the two important parameters $C_c$ and $\Theta_{CW}$ have been deduced reliably by the fitting.

The magnitude of the Curie constant means that every Mn ion possesses an effective paramagnetic moment of $p_{eff}$ = 4.93. This value is nearly equal to the spin-only value of $g\sqrt{S(S+1)}$ = 4.90 for $g$ = 2 and $S$ = 2, where $g$ is the Lande $g$ factor. In other words, the $g$ factor is 2.01, which is a typical value for the Mn$^{3+}$ ion with four $d$ electrons in the high-spin configuration. Therefore, 3$d$ electrons are localized thoroughly at the Mn site in Ag$_2$MnO$_2$.

On the other hand, the large, negative Curie-Weiss temperature indicates a significant AF interaction between Mn spins. This is in contrast to that in the case of Ag$_2$NiO$_2$, where $\Theta_{CW}$ values are 10 and -30 K above and below the orbital ordering temperature, respectively.[25] It is known for related compounds having similar NiO$_2$ layers that magnetic interactions are often ferromagnetic because of the nearly 90 degree Ni-O-Ni superexchange paths: $\Theta_{CW}$ = 20 - 30 K and 36 K for NaNiO$_2$ and LiNiO$_2$, respectively.[29, 30] Thus, the large AF value for the present manganite is rather exceptional. Probably, complicated superexchange interactions through bridging oxygens or direct exchange couplings between $t_{2g}$ orbitals should be taken into account. Moreover, the distortion of MnO$_2$ layers that modifies the Mn-O-Mn bond angle from 90 deg may play a role in the large AF interactions.

It is not clear in the present study how large the interplane coupling is, although predominant magnetic interactions should lie in the triangular plane. In terms of the crystal structure, the interplane coupling may be small because of the large separation of ~0.8 nm between the planes compared with the short Mn-Mn distance of ~0.3 nm in the plane. Also, note that there are no superexchange pathways connecting the planes across the $Ag_2$ layer. However, an additional RKKY interaction through conduction electrons may exist, although it is difficult to estimate how large it is. Here, we stress that $Ag_2MnO_2$ does not undergo a conventional LRO down to 2 K, as described below; thus, the two dimensionality of magnetic interactions may be reasonably good.

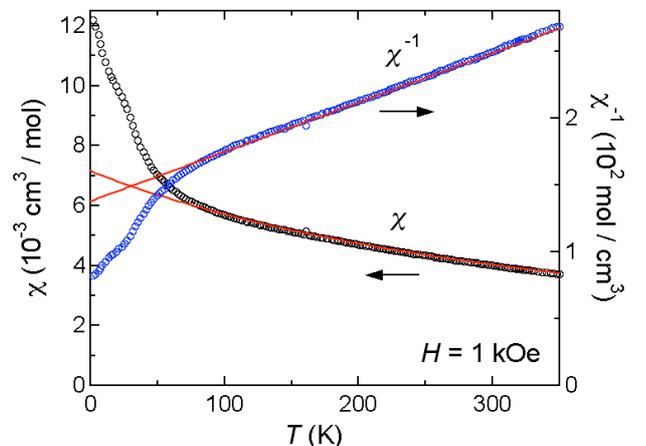

Fig. 3. Magnetic susceptibility $\chi$ and its inverse measured on cooling for polycrystalline sample of Ag$_2$MnO$_2$ at magnetic field of 1 kOe. The solid line in the inverse $\chi$ plot is a linear fit, and that in the $\chi$ plot is a fit to the form of $\chi = \chi_0 + \chi_{CW}$.

According to the molecular field theory, the magnitude of magnetic interactions between two nearest-neighbor Mn spins is estimated to be 18 K; $|\Theta_{CW}| = 2zS(S+1)J / 3k_B = 24J/k_B$, where $z = 6$ is the number of nearest neighbors, assuming that $H = 2J\sum_{\langle i,j \rangle} S_i \cdot S_j$. This estimate may be justified for such a classical spin system. In fact, in the case of the hexagonal antifferromagnet $YMnO_3$ containing similar $Mn^{3+}$ ions, $\Theta_{CW}$ is -705 K and the average $J$ determined by inelastic neutron scattering experiments is 30 K: their ratio is 23.[20]

On further cooling, the magnetic susceptibility deviates from the high-temperature CW behavior below ~100 K and finally shows a shoulder at ~25 K. We will rediscuss these important features later.

*3.3 Phase transition at 80 K*

Figure 4 shows the specific heat of $Ag_2MnO_2$ in a wide temperature range. At first glance, there is no discernible anomaly in the $T$ dependence. Also plotted in Fig. 4a is the specific heat of $Ag_2NiO_2$, where two anomalies are observed at $T_s = 260$ K and $T_N = 56$ K. Except for these enhancements, the two curves overlap well in a wide temperature range, suggesting similar lattice contributions, as expected from the similarity in crystal structure. When the two curves are compared carefully, however, it is noticed that the Mn curve shifts upward slightly below ~80 K compared with the Ni curve, as is apparent in the enlarged figure of the inset.

In order to estimate the background line, we fitted the Mn data in the whole temperature range except at 35-100 K, by incorporating three Debye-type and one Einstein-type contributions; $C_b = \gamma T + xC_{D1} + yC_{D2} + zC_{D3} + (5 - x - y - z)C_E$, where $C_{D1}$, $C_{D2}$ and $C_{D3}$ are from Debye-type phonons characterized by Debye temperatures $\Theta_{D1}$, $\Theta_{D2}$ and $\Theta_{D3}$ and $C_E$ is from an Einstein-type phonon with an Einstein temperature $\Theta_E$. The first term comes from conduction electrons with $\gamma = 20.6$ mJ $K^{-2}$ $mol^{-1}$, as determined in Fig. 2. The result of fitting is fairly good, as shown by the solid line in Fig. 4a, yielding $\Theta_{D1} = 172(4)$ K, $\Theta_{D2} = 460(20)$ K, $\Theta_{D3} = 1020(20)$ K, $x = 1.69(9)$, $y = 1.73(4)$, and $z = 1.49(6)$. The lowest Debye temperature is determined uniquely by the initial rise in specific heat from $T = 0$ K, and the other two contributions are necessary to reproduce the high temperature part. The contribution of an Einstein phonon is relatively small, i.e., $5 - x - y - z = 0.08$, but its incorporation has considerably improved the fit at low temperatures. It is considered in the first-order approximation that the Debye phonon with the lowest Debye temperature may come from the $Ag_2$ layers of metal bonding, and the other Debye phonons from the $MnO_2$ layers of ionic bonding. Since such a virtual separation may not be a good approximation in the low-temperature limit, one expects a unified Debye phonon with an intermediate $\Theta_D$ like 236 K, as shown in Fig. 2.

Then, residual contributions have been obtained by subtracting the above background and are shown in Fig. 4b. The small anomaly is now evident as a sudden jump at 80 K, followed by a gradual decrease on cooling. It is concluded from the shape that there is a second-order phase transition at $T_c = 80$ K. We prepared and examined several samples and found that some samples exhibited a broadened transition possibly due to poor sample quality (we do not know how poor they are at the moment). Note, however, that the anomaly was always detected at 80 K. The entropy consumed at the transition is evaluated by integrating $C/T$ to be approximately 0.36 J $K^{-1}$ $mol^{-1}$, which is small compared with the maximum total entropy for $S = 2$ spins; $R\ln(2S + 1) = 13.38$ J $K^{-1}$ $mol^{-1}$, where $R = 8.314$ J $K^{-1}$ $mol^{-1}$ is the gas constant. The rest of the entropy may be distributed in a wide temperature range and is submerged in the background.

On the other hand, the same lattice contribution is used for the Ni compound, giving a magnetic contribution associated with the AF transition at $T_N = 56$ K, as shown in Fig. 4b. The entropy change for the magnetic transition in the Ni compound is 1.24 J $K^{-1}$ $mol^{-1}$, about 20% of $R\ln2 = 5.76$ J $K^{-1}$ $mol^{-1}$ for $S = 1/2$. Moreover, the entropy change due to structural transition is 4.81 J $K^{-1}$ $mol^{-1}$, close to that for $R\ln2$ expected for the $e_g$ orbital degree of freedom, which may justify the above estimation of the lattice term for the two compounds.

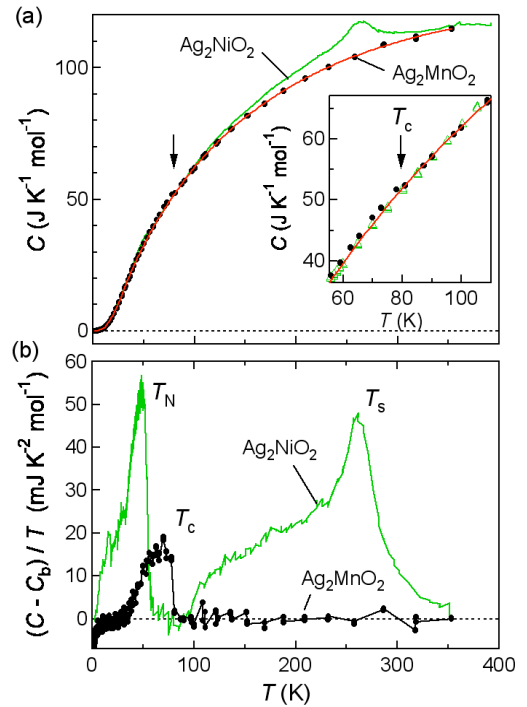

Fig. 4. (a) Specific heats $C$ of $Ag_2MnO_2$ (circle) and $Ag_2NiO_2$ (line or open triangle). The solid line near the circles is a fit for estimating the background. The inset expands the data at around 80 K. (b) Specific heats after subtraction of the background for the two compounds.

Going back to the magnetic susceptibility in Fig. 3, there is no corresponding anomaly at $T_c = 80$ K, only a smooth increase in $\chi$. However, the deviation from the CW law begins at approximately $T_c$. In addition, we found a significant magnetic field dependence in $\chi$ below $T_c$, as shown in Fig. 5a: two data sets measured on cooling at magnetic fields of 1 and 70 kOe completely overlap above $T_c$, while they are separate below $T_c$. A change of the field dependence across $T_c$ is also observed in the isothermal $M$-$H$ measurements shown in Fig. 5b: the $M$-$H$ curves measured at 200 K and 100 K are perfectly linear, namely, paramagnetic, while those at 50, 20 and 2 K show concave-downward curvatures, indicating a weak ferromagnetic correlation. Therefore, it is concluded that the phase transition at $T_c$ is accompanied by a certain change in magnetism. Probably, a short-range AF order incorporating a weak ferromagnetism may develop below $T_c$.

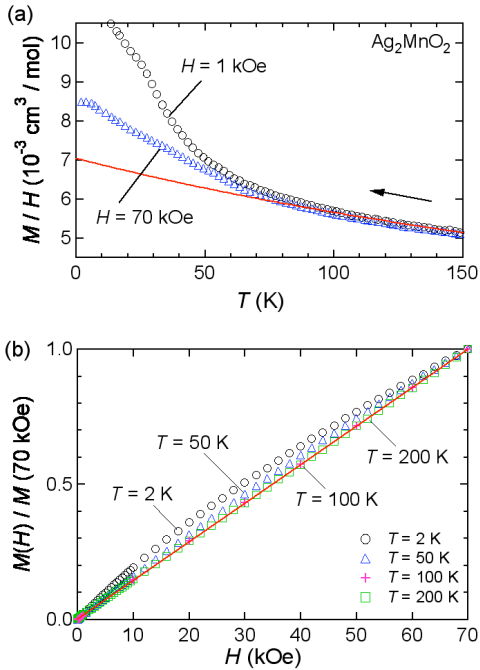

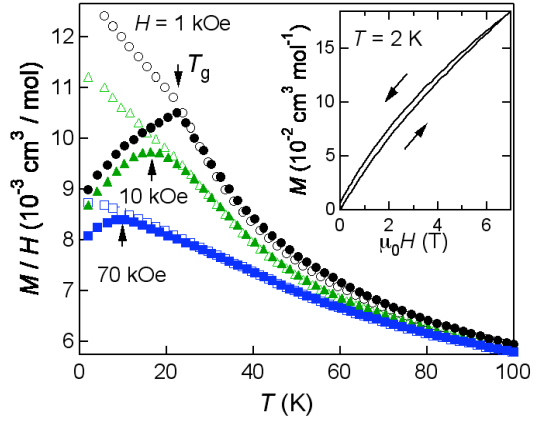

Fig. 6. Magnetic susceptibility measured on heating after zero-field cooling (solid marks) and on subsequently cooling (open marks) at magnetic fields of 1, 10 and 70 kOe. The inset shows the $M$-$H$ curves measured at $T = 2$ K on increasing and decreasing fields.

Fig. 5. (a) Magnetic susceptibility measured on cooling at magnetic fields of 1 and 70 kOe. (b) Magnetic field dependence of magnetization normalized to the data at $H = 70$ kOe measured at various temperatures with increasing field after zero-field cooling. The solid line is a linear fit to the data

*3.4 Spin freezing below 22 K*

Finally, we describe the low-temperature anomaly found in $\chi$. Three experiments carried out at magnetic fields of 1, 10 and 70 kOe are shown in Fig. 6. Commonly, the first run measured on heating from 2 K after zero-field-cooling (ZFC) exhibits a cusp or a broad peak, and the second, subsequent run on cooling (FC) separates from the ZFC curve near the peak temperature and shows a further increases on cooling. The peak temperature $T_g$ decreases with increasing field. Moreover, the clear cusp observed in the ZFC curve at 1 kOe at $T_g = 22$ K becomes smaller and broadens with increasing field. Thus, there is a large thermal hysteresis that tends to be suppressed under large magnetic fields. Probably related to this, a distinct hysteresis between two $M - H$ curves measured on increasing and de-creasing field is detected at $T = 2$ K, resulting in a small remanent magnetization at zero field, as shown in the inset in Fig. 6. In contrast, there was no hysteresis for the data taken at 20 and 50 K.

The observed cusp in dc $\chi$ with a distinct thermal hysteresis is reminiscent of a spin or cluster glass transition. It should not be ascribed to a conventional LRO, because the corresponding anomaly is missing in specific heat. In order to determine the possibility of spin-freezing transition, we measured ac susceptibility, $\chi' - i\chi''$, at various frequencies of the ac field. Figure 7a shows a typical data measured on cooling at a field amplitude $h = 3$ Oe and a frequency $f = 5.1$

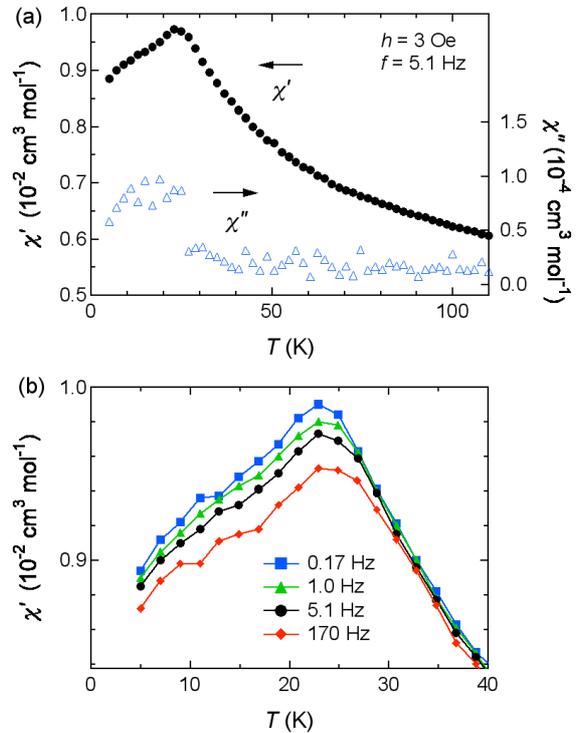

Fig. 7. ac susceptibility of $Ag_2MnO_2$ measured on cooling at a magnetic field amplitude $h = 3$ Oe and at a frequency $f = 5.1$ Hz (a) and various frequencies $f = 0.17, 1.0, 5.1, 170$ Hz (b).

Hz. The real part exhibits a broad peak at ~23 K, which is close to $T_g$ = 22 K found in dc $\chi$, while the imaginary part increases suddenly at ~ 25 K. Figure 7b shows four sets of $\chi'$ data at low temperature measured at various frequencies, where no frequency dependence is observed at high temperatures above $T_g$, but is clearly observed below around $T_g$, as observed in typical spin glass systems.[31] This means that a certain freezing of spins takes place below $T_g$, not a conventional LRO.

Nevertheless, the ground state of spins in $Ag_2MnO_2$ should not be considered as a conventional spin glass, because there is no inherent disorder in the system. Recently, frustration-induced spin-glass-like behavior has been observed in the 3D pyrochlore lattice of some pyrochlore molybdates.[32] A typical compound that exhibits a spin-freezing transition at 22 K is $Y_2Mo_2O_7$.[33] The $T$-, $H(h)$- and $f$-dependences of dc and ac susceptibilities reported for $Y_2Mo_2O_7$[34] are quite similar to those that have been observed for $Ag_2MnO_2$. Thus, a similar frustration-induced spin freezing must take place in this 2D spin system. It is plausible that the SRO of spins develops below $T_c$, but cannot develop into LRO owing to frustration. Instead, small clusters with a spin order accompanied by a weak ferromagnetic moment may freeze randomly below $T_g$.

## 4. Discussion

Let us summarize first how the $Mn^{3+}$ spins of $Ag_2MnO_2$ behave on cooling. At high temperatures above 100 K, they are paramagnetic so that $\chi$ perfectly obeys the Curie-Weiss law with a large AF $\Theta_{CW}$ of -430 K, which demonstrates a deviation from the classical AF magnet described by a mean field theory: neither LRO near $T = |\Theta_{CW}|$ nor even SRO develops down to almost a quarter of $|\Theta_{CW}|$. This is obviously due to geometrical frustration as well as to a good two dimensionality, as generally observed for frustrated magnets.[2] At $T_c$ = 80 K, a certain order sets in, accompanied by the SRO of spins growing below $T_c$. Finally, at $T_g$ = 22 K, the spins may nearly undergo LRO within a cluster, and the clusters freeze randomly with each other, resulting in a glassy state. The presence of weak ferromagnetic correlations may be attributed to the Dzyaloshinsky-Moriya interaction in the $MnO_2$ layer, as often observed in frustrated spin systems.

The transition at $T_c$ = 80 K is of the second order and is characterized by a clear sign in specific heat, with no cusp in magnetic susceptibility, as reproduced in Fig. 8. The reverse case is often found in experiments: no anomaly is detected in $C$ but in $\chi$, implying that the anomaly is not due to a phase transition in bulk nature. This is because $\chi$ tends to be affected significantly by a small amount of impurity or a minor region in a sample, particularly in the case of antiferromagnets. Thus, the present case is quite unusual. One possibility is that the transition is purely of structural origin. To confirm this, we performed low-temperature XRD experiments and found no evidence of a structural transition, that is, no detectable changes in the powder XRD pattern down to 4 K. It is difficult, however, to exclude the possibility of such transition completely. We can just say that the transition at $T_c$ is not accompanied by a large change in crystal structure. In the case of $Ag_2NiO_2$, the trigonal structure is deformed actually to a monoclinic structure below 260 K to lift the orbital degeneracy.[25, 26] In contrast, the structure of $Ag_2MnO_2$ is already distorted at room temperature; thus, one may not expect a further change at lower temperatures driven by the Jahn-Teller mechanism.

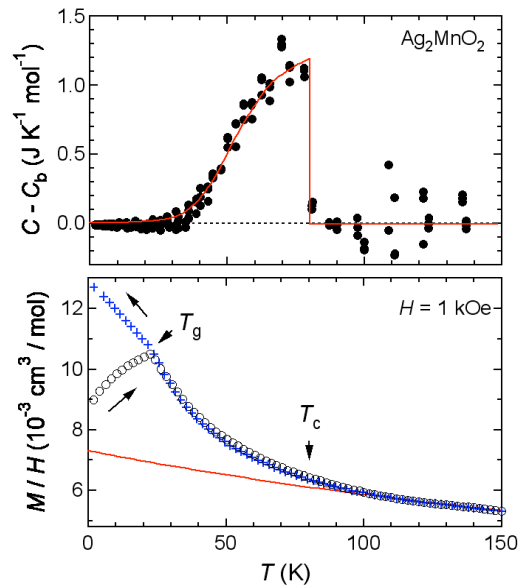

Fig. 8. Specific heat (top) and magnetic susceptibility (bottom) showing presence of second-order phase transition at $T_c$ = 80 K and absence of corresponding anomaly, respectively. The solid line in the specific heat data serves as a visual guide.

On the other hand, the two experimental facts on magnetic properties that a deviation from CW behavior takes place at approximately $T_c$ and that a weak ferromagnetic correlation develops below $T_c$ may indicate a certain magnetic transition. In addition, very recent µSR experiments on our sample by Sugiyama *et al.* have revealed that rapidly fluctuating moments appear below ~80 K probably owing to the formation of SRO, while a "static" order is completed below ~30 K.[35] This strongly supports the magnetic phase transition at $T_c$ = 80 K in $Ag_2MnO_2$.

The most interesting scenario is to assume a phase transition associated with chirality, which was predicted theoretically for the *XY* spin system on a triangular lattice, but has never been observed experimentally. The chirality of spins is defined for each elementary triangle as

$$\kappa = \frac{2}{3\sqrt{3}}(\mathbf{S}_2 \times \mathbf{S}_1 + \mathbf{S}_3 \times \mathbf{S}_2 + \mathbf{S}_1 \times \mathbf{S}_3),$$

where spins are numbered counterclockwise, as shown in Fig. 9. In the classical *XY* spin model, the chirality vector is always perpendicular to the plane and becomes a scalar variable analogous to an Ising spin expressed as

$$\kappa = \frac{2}{3\sqrt{3}}(sin(\theta_2 - \theta_1) + sin(\theta_3 - \theta_2) + sin(\theta_1 - \theta_3)),$$

where $\theta_i$ is the angle of the $i$-th spin. For an ideal 120° structure, $\kappa$ takes +1 or -1 alternately for neighboring triangles, as shown in the bottom left of Fig. 9.

As mentioned in the introduction, MS found that $\kappa$ vanishes steeply at $k_B T_{ch} / J$ = 0.513, which is to be ascribed to an order-disorder type, second-order phase transition of chirality:[14] Since the center of triangles form a honeycomb lattice that is a bipartite lattice, the chirality itself is not frustrated and can undergo LRO of the Ising type. Interestingly, specific heat shows a logarithmic divergence at $T_{ch}$, while magnetic susceptibility is nearly $T$-independent, as observed in the present case of $Ag_2MnO_2$. The weak response in magnetic susceptibility at $T_{ch}$ means that spins are still fluctuating below $T_{ch}$ even after the chirality transition sets in, as intuitively illustrated in Fig. 9. The figure reproduces part of the snapshot of the spin and chirality configurations taken at a temperature near $T_{ch}$ given by MS.[14] Note that two triangles having positive and negative $\kappa$ values ($|\kappa| < 1$) are aligned perfectly in a staggered way, while spins on each of the three sublattices change their directions from place to place, that is, are still fluctuating in SRO.

At higher temperatures, spins are randomly oriented in the plane so that the sign of chirality is also distributed randomly in a snapshot. Below $T_{ch}$, on the other hand, spin correlations begin to grow toward the 120° structure; as a result, the $\kappa$ of one triangle should be opposite in sign to that of the adjacent three triangles. Since this can cover all the triangles consistently, the LRO of $\kappa$ is realized. As temperature decreases further, the SRO of spins becomes LRO at $T_{KT}$, that is, the spin-correlation length becomes infinite. Therefore, a unique state is found in the temperature window between $T_{ch}$ and $T_{KT}$.

Entropy released at the chirality transition may be smaller than the total value, which should correspond to the chirality degree of freedom. The small entropy change at $T_c$ found in $Ag_2MnO_2$ is to be compared with a theoretical value that may not be available at the moment. The rest of the spin entropy may be distributed over a wide temperature range below $T_c$.

Concerning the critical temperature, we have obtained $k_B T_c /J = 80 / 18 = 4.4$. MS found $k_B T_{ch} /J' = 0.513$ on the basis of the Hamiltonian $H = J' \sum_{\langle i,j \rangle} S_i \cdot S_j, S_i = (\cos\theta_i, \sin\theta_i)$.[14] Lee et al. also obtained a similar $T_{ch}$ value: $k_B T_{ch} /J' = 0.510(5)$.[16] To translate their $T_{ch}$ to ours, $J'$ should be replaced with $2JS^2$, which results in $k_B T_{ch} /J = 4.1$. Therefore, our $T_c$ is reasonably close to the theoretical $T_{ch}$ predicted for chirality transition.

MS and other researchers pointed out that $XY$ spins on a triangular lattice also exhibit a KT transition at $k_B T_{KT} / J = 0.502$ or $0.505(5)$.[14, 16] Because of the proximity of the two predicted temperatures, however, it would be difficult to detect the two transitions separately in experiments. What we have observed for $Ag_2MnO_2$ is that only SRO with weak ferromagnetic correlations develops below $T_c$, followed by a glasslike transition at $T_g$. Since spin correlations must be enhanced critically at $T_g$, it is plausible that glass transition replaces KT transition in $Ag_2MnO_2$. Possibly, only $T_{KT}$ is reduced to a temperature lower than the theoretically expected temperature and merges with $T_g$ for some reasons. We suspect that a higher-order term in a spin Hamiltonian is required to understand the separation of the two temperatures.

It is not clear in the present study that the $Mn^{3+}$ spin in $Ag_2MnO_2$ is considered as an $XY$ spin. However, similar $Mn^{3+}$ spins have been assumed to be $XY$ spins in previous studies of $YMnO_3$.[18-20] On the other hand, $S = 2$ is the largest anisotropic spin among $d$ electron systems. Thus, we expect that $Mn^{3+}$ spins on a triangular lattice will provide us with the most adequate system for realizing a classical $XY$ spin model.

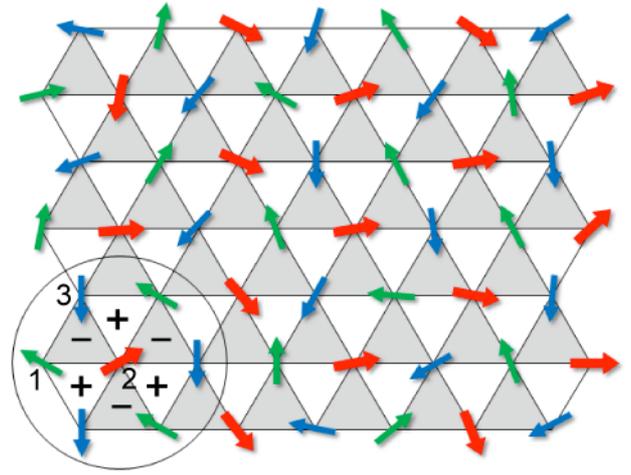

Fig. 9. Schematic representation of chirality order that reproduces part of Fig. 12 of MS paper.[14] It is a snapshot illustrating a characteristic feature of the spin and chirality configurations calculated by Monte Carlo simulations at a temperature near the chirality transition temperature $T_{ch}$. The spin on each lattice point is represented by an arrow. The ideal 120° structure expected at $T = 0$ K is depicted in a circle on the bottom left, which contains three sets of spins forming a 120° angle to each other. The marks + and - on triangles indicate $\kappa = +1$ and -1, respectively. $\kappa$ can take intermediate values at an elevated temperature. For example, as depicted here outside the circle, upward shaded triangles have negative $\kappa$ values of $-1 < \kappa < -0.3$, while downward open ones have positive $\kappa$ values of $0.3 < \kappa < 1$. These critical values of $\pm 0.3$ have been chosen so as to attain such a perfect staggered order of chirality as drawn here, and should depend on the temperature selected. In spite of the fact that the chirality has undergone perfect LRO, spins on each of the three sublattices are not parallel to each other and still markedly fluctuate in SRO.

Nevertheless, no real spins can completely lose their $z$-axis component. Thus, one has to always consider an intermediate situation between the $XY$ and Heisenberg Hamiltonians. In the pure Hiesenberg model, Kawamura and Miyashita found a KT transition associated with $Z_2$ vortices at $k_B T_{KT} / J' = 0.33$, which is ~30% lower than that for the $XY$ model.[36] It seems that there are no other transitions in the isotropic case. Possibly, the chirality degree of freedom may not be independent in such a case. Note for the Heisenberg model that specific heat shows a broad peak, not a logarithmic divergence, at $T_{KT}$, while uniform susceptibility exhibits a much smaller one, as in the case of the chirality transition for the $XY$ model. It is intriguing how these transitions in the two extreme models are linked to each other.[37, 38] Experimentally, it is important to clarify where $Ag_2MnO_2$ exists between the $XY$ and Heisenberg models.

## 5. Conclusions

In summary, we have studied the magnetic properties of the novel metallic compound $Ag_2MnO_2$ that contains $Mn^{3+}$ spins with $S = 2$ on the triangular lattice. We found a unique phase transition at $T_c = 80$ K, where specific heat shows an anomaly but magnetic susceptibility does not. We believe that it is ascribed to the order-disorder transition of chirality theoretically predicted for the classical $XY$ spin system on a triangular lattice. To clarify the nature of the transition, further study hopefully using a single crystal is required.

## Acknowledgements

We are grateful to M. Ichihara for the electron diffraction analysis, to Y. Kiuchi for the chemical analysis and to S. Miyashita and M. Ogata for helpful discussions. This work was supported by a Grant-in-Aid for Scientific Research on Priority Areas "Novel States of Matter Induced by Frustration" (No. 19052003).